\documentclass[12pt]{iopart}
\usepackage{iopams}
\usepackage{graphicx}
\usepackage{color}

\begin{document}

\def\bra#1{\langle #1 |}
\def\ket#1{| #1 \rangle}
\newcommand{\REV}[1]{{\color{red}#1}}
\newcommand{\BLUE}[1]{\textbf{\color{blue}#1}}
\newcommand{\GREEN}[1]{\textbf{\color{green}#1}}
\newcommand{\rev}[1]{{\color{red}#1}}
\newcommand{\blue}[1]{{\color{blue}#1}}

\title{Unearthing wave-function renormalization effects in the time evolution of a Bose-Einstein condensate}

\author{Paolo Facchi$^1$, Saverio Pascazio$^1$, Francesco V.\ Pepe$^1$,
Ennio Arimondo$^{2,3}$, Donatella Ciampini$^{2,3}$ \& Oliver Morsch$^3$}

\address{$^1$Dipartimento di Fisica and MECENAS, Universit\`a di Bari, I-70126 Bari, Italy \\ 
and INFN, Sezione di Bari, I-70126 Bari, Italy}
\address{$^2$CNISM-Pisa, Dipartimento di Fisica, Universit\`{a} di Pisa, Lgo Pontecorvo 3, 56127 Pisa, Italy}
\address{$^3$INO-CNR, Dipartimento di Fisica, Universit\`{a} di Pisa, Lgo Pontecorvo 3, I-56127 Pisa,Italy}

\begin{abstract}
We study the time evolution of a Bose-Einstein condensate in an accelerated optical lattice. When the condensate has a narrow quasimomentum distribution and the optical lattice is shallow, the survival probability in the ground band exhibits a steplike structure. In this regime we establish a connection between the wave-function renormalization parameter $Z$ and the phenomenon of resonantly enhanced tunneling.
\end{abstract}

\pacs{03.75.Lm, 03.65.Xp, 11.10.Gh}

%\noindent{\it Keywords}: BECs, Thomas-Fermi approximation, Boson mixtures.

%\maketitle

\section{Introduction}

Renormalization effects are among the most subtle and interesting in quantum field theory \cite{textbooks} and go to the heart of our understanding of fundamental interactions.
Free field operators create and annihilate single particles with unit probability. The presence of interactions makes things much more complicated and requires powerful analytical tools, in order to compute the propagator and the $S$-matrix through the self-energy function and guarantee probability conservation.
In the K\"all\'en-Lehmann representation \cite{KL}, probability conservation is enforced via wave function renormalization, that consists in a rescaling of quantum fields by a factor $\sqrt{Z}$ to take into account the effects of interactions. 

The propagator yields exponential decay (Fermi ``golden" rule) due to the contribution of a pole on the second Riemann sheet in the complex energy plane \cite{temprevi}, but gets also a contribution of order (coupling constant)$^2$ from a contour integration in the complex energy plane, that 
modifies the exponential decay law both at short and long times, yielding the characteristic quadratic and power-law behaviors. 
The quantity $\sqrt{Z}$ is the modulus of the residue of a pole of the propagator in the second Riemann sheet in the
complex energy plane and is different from unity when the latter is computed beyond leading order in the coupling constant. For a stable state one has $Z<1$ (due to probability conservation in the
K\"all\'en-Lehmann representation), but for an unstable state $Z$ is unconstrained and can be $>1$ \cite{Brown}. This is a common feature of quantum decay processes arising from the coupling of a discrete level to a continuum. In such a case the discrete level moves away from the real axis and acquires an imaginary part, which accounts for its decay rate.  Since $Z$ is the square of the overlap between the initial state and this non-normalizable complex-energy vector, known as ``Gamow state'' \cite{Gamow,Bohm,Gadella}, one can have $Z>1$. 

\begin{figure}
\begin{center}
\includegraphics[width=0.5\textwidth]{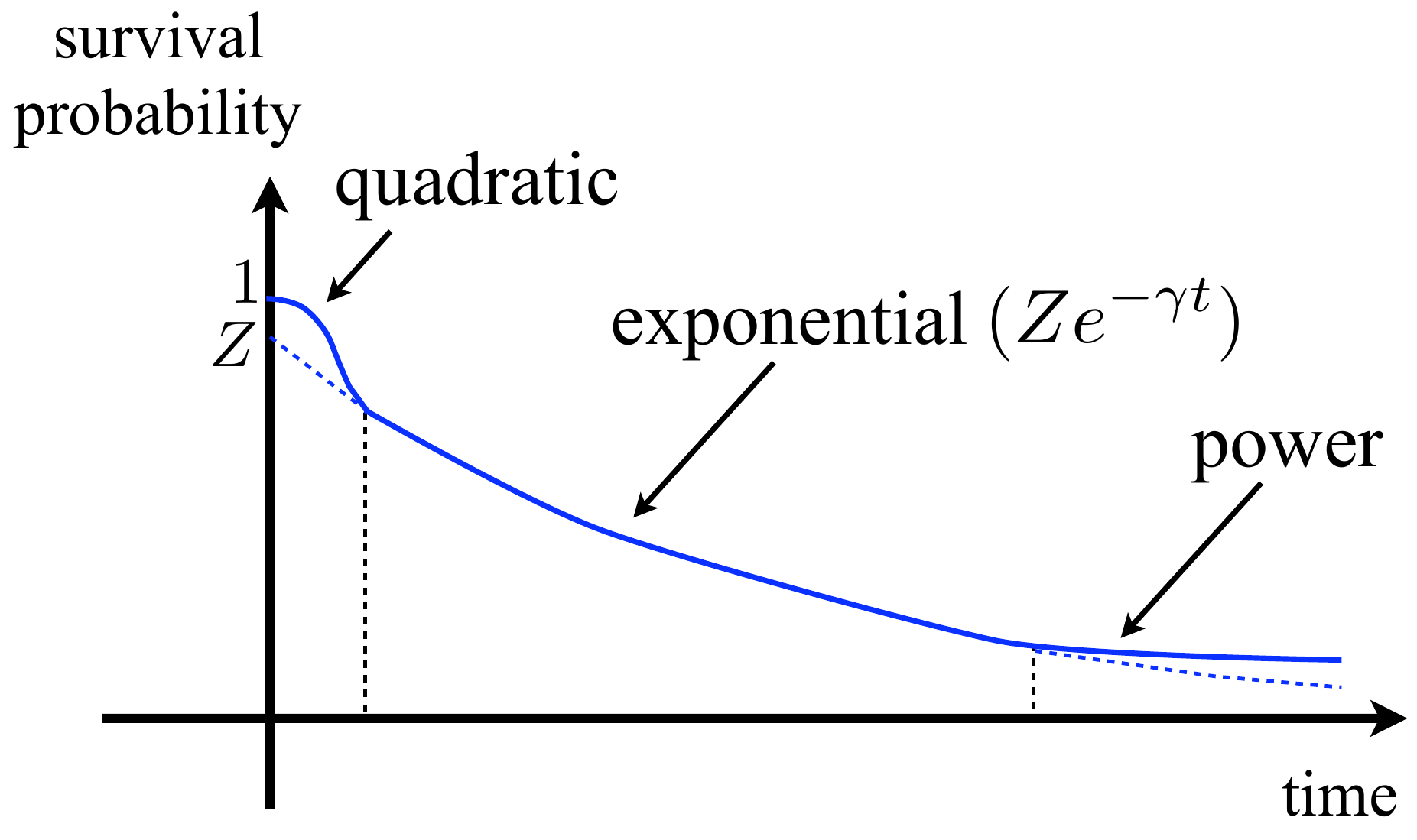}
\end{center}
\caption{Survival probability of an unstable system. The exponential decay law $P(t) \simeq P_{\rm{Z}}(t)= Z e^{-\gamma t}$ is only valid at intermediate times and its extrapolation at $t=0$ yields wave function renormalization $Z$. In general $Z \neq 1$.}
\label{fig:tevol}
\end{figure}

A typical time evolution of the survival probability of an unstable system in its initial state is displayed in Fig.~\ref{fig:tevol}. The exponential law is approximately valid in an intermediate time region. It is preceded by a quadratic decay (Zeno region) and superseded by a power law. Notice that the extrapolation of the exponential to $t=0$ yields a value (wave function renormalization $Z$) that is in general $\neq 1$. We shall write 
\begin{equation}
\label{genevol}
P(t) = Z e^{-\gamma t} + \textrm{additional contributions},
\end{equation}
where the additional contributions are typically second order in the coupling (to a continuum of states to which the system decays) and are important at short and long times. At intermediate times the exponential is assumed to dominate and characterize the evolution
\begin{equation}
\label{evolZ}
P(t) \simeq P_{\rm{Z}}(t)= Z e^{-\gamma t} ,
\end{equation}
with $\gamma$ given by the Fermi golden rule.

The short-time quadratic (and hence non-exponential) decay was first experimentally observed by the group led by Mark Raizen in Texas \cite{Wilkinson97, raizen}. In this pioneering work focus was on the occurrence of the quantum Zeno effect and its inverse \cite{FP,KK} due to repeated measurements.
In this article we shall focus on the same physical system used in \cite{Wilkinson97,raizen}, but under very different physical conditions, both in terms of initial state and physical parameters \cite{LP}. This will enable us to unearth wave-function renormalization effects. 

\begin{figure}
\begin{center}
\includegraphics[width=0.5\textwidth]{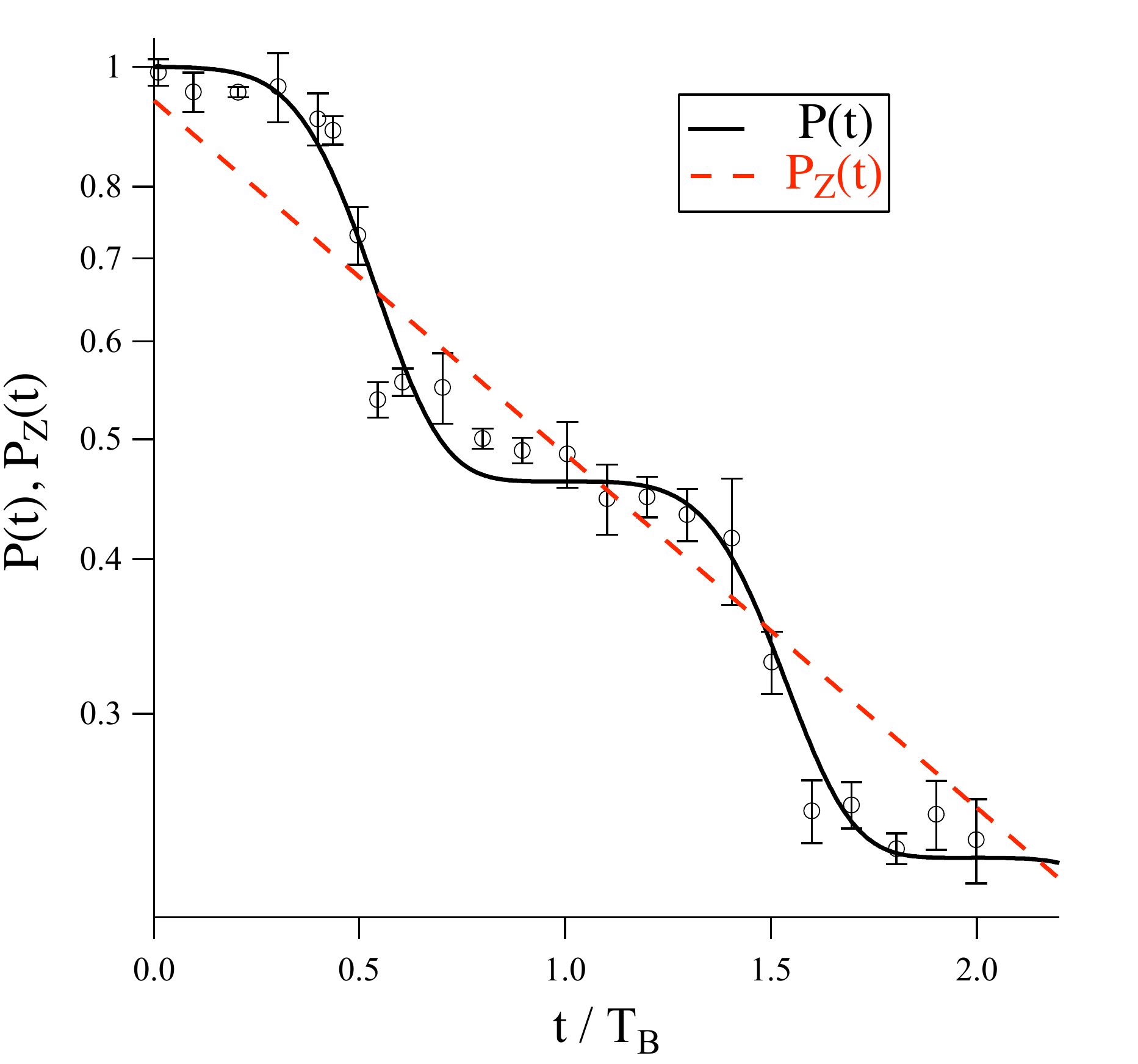}
\end{center}
\caption{Survival probability $P(t)$ of the wave packet of the atomic cloud in the lowest energy band of the accelerated optical lattice.
The numerical solution of the Schr\"odinger equation is the full line. The experimental data (open circles) are the same as in Ref.\ \cite{LP} and the (red) dashed line is the exponential fit based on $P_{\rm Z}(t)$ in Eq.\ (\ref{expfit}), whose crossing  with the $y$ axis yields the value of $Z$. The slope of the exponential decay yields the decay rate $\gamma$ (Fermi Golden rule). In this plot $V_0 = 1 E_{\textrm{rec}}$, $F_0= 0.383$, yielding $Z<1$. }
\label{SurvivalProbability}
\end{figure}

Let us briefly sketch the main ideas to be explored in this note. We shall investigate the time evolution of a wave packet (a ``cloud" of  atoms in a Bose-Einstein condensate) in a tilted optical lattice. The wave packet, initially in the lowest energy band, is narrow in momentum space and feels a constant acceleration due to the lattice tilt. When it reaches the edge of the band, it can make a Landau-Zener transition \cite{LZ} to the upper band. The survival probability in the lowest energy band is therefore characterized by plateaus (when the packet evolves in the band) and sharp transition regions (when the packet loses probability to the upper band). 
A typical example is displayed in Fig.\ \ref{SurvivalProbability}. One clearly sees that in practice, the dynamics can be rather different from the idealized situation depicted in Fig.\ \ref{fig:tevol}.
The overall process is involved and the quantum dynamics depends on a plethora of factors, ranging from probability losses to phase coherence effects. We shall see that Landau-Zener transitions \cite{LZ} and St\"uckelberg's oscillations \cite{stuckosc} are two facets of this complex evolution and that by scrutinizing the features of the survival probability of the wave function that collectively describes a cold atomic cloud, one can extract crucial information on wave-function renormalization effects.

In this article we shall endeavor to give an elementary introduction to this problem and discuss in more detail the model adopted in Ref.\ \cite{LP} to study these effects. We shall also scrutinize the convergence of the decay rate and wave-function renormalization parameter to their asymptotic values.
The main idea will be to use Landau-Zener transitions as a benchmark for the study of wave-function renormalization effects.

\section{Landau-Zener tunneling}

A Landau-Zener (LZ) transition takes place in a two-level system
with a time-dependent Hamiltonian $H_{\mathrm{LZ}}(t)$ whose
spectrum, as a function of time, is characterized by the presence
of an avoided crossing 
\cite{LZ,stuckosc,Majorana,Jones}. In the
time-independent, {\it diabatic basis} $\{\ket{1},\ket{2}\}$, the expectation values of the Hamiltonian are linear in time
\begin{equation}
\bra{1} H_{\mathrm{LZ}}(t) \ket{1} = -\bra{2}
H_{\mathrm{LZ}}(t) \ket{2} = -\alpha t,  \qquad (\alpha >0)
\end{equation}
while the off-diagonal matrix elements are
constant in time and can be assumed to be real by properly
adjusting the phase of the basis states
\begin{equation}
\bra{1} H_{\mathrm{LZ}}(t) \ket{2} = \bra{2}
H_{\mathrm{LZ}}(t) \ket{1} = \delta.
\end{equation}
The state $\ket{\psi(t)}$ of the system is expanded as
\begin{equation}
\ket{\psi(t)} = a_1(t) \ket{1} + a_2(t) \ket{2}
\end{equation}
and the Schr\"{o}dinger equation reads
\begin{equation}
i\hbar\frac{d}{dt} \left( \begin{array}{c} a_1 \\ a_2 \end{array}
\right)= \left( \begin{array}{cc} -\alpha t & \delta \\ \delta &
\alpha t \end{array} \right) \left( \begin{array}{c} a_1 \\ a_2
\end{array} \right).
\end{equation}
The time-dependent eigenbasis of $H_{\mathrm{LZ}}$ is called {\it adiabatic basis} and can be obtained by a rotation
\begin{eqnarray}
\left( \begin{array}{c} \ket{1(t)} \\ \ket{2(t)} \end{array}
\right)= \left(
\begin{array}{rr} \cos\theta(t) & -\sin\theta(t) \\ \sin\theta(t) &
\cos\theta(t)
\end{array} \right) \left( \begin{array}{c} \ket{1} \\ \ket{2}
\end{array} \right), 
\end{eqnarray}
with
$\theta(t)= \arctan (\alpha t/\delta)/2$.
The adiabatic vectors satisfy
\begin{eqnarray}
H_{\mathrm{LZ}}(t) \ket{1(t)} = - \Omega(t) \ket{1(t)} , \quad 
H_{\mathrm{LZ}}(t) \ket{2(t)} = \Omega(t) \ket{2(t)} ,  
\end{eqnarray}
with 
\begin{equation}
\Omega(t)= \sqrt{(\alpha t)^2+\delta^2}.
\end{equation} 
The degeneration of the mean energy of the diabatic states at $t=0$ is
reflected in an avoided energy level crossing in the adiabatic basis, with the
two states reaching the minimum energy distance at $t=0$. By setting
\begin{equation}
\ket{\psi(t)} = b_1(t) \ket{1(t)} + b_2(t) \ket{2(t)},
\end{equation}
from
\begin{equation}
\frac{d}{dt}  \ket{1(t)} = - \dot{\theta}(t) \ket{2(t)},
\qquad \frac{d}{dt}  \ket{2(t)} = + \dot{\theta}(t)
\ket{1(t)}
\end{equation}
we get 
\begin{equation}
i\hbar\frac{d}{dt} \left( \begin{array}{c} b_1 \\ b_2 \end{array}
\right)= \left( \begin{array}{cc} -\Omega(t) & -i \hbar \dot{\theta}(t) \\
i \hbar \dot{\theta}(t) & \Omega(t)
\end{array} \right) \left(\begin{array}{c} b_1 \\ b_2
\end{array} \right) .
\end{equation}
Therefore, in the adiabatic basis, the evolution of the system is
governed by the Hamiltonian
\begin{equation}
\label{Heff}
H_{\rm{ad}}(t) = - \Omega(t) \sigma_z + \Gamma(t) \sigma_y, \qquad
\Gamma(t) = \hbar \dot{\theta}(t) 
= \frac{\hbar \alpha \delta}{2 \Omega^2
(t)},
\end{equation}
with $\sigma_y$ and $\sigma_z$ %respectively the second and third
the Pauli matrices. %and
%\begin{eqnarray}
%\Gamma(t) \equiv \hbar \dot{\theta}(t) = \frac{\hbar}{2} \frac{\alpha \delta}
%{(\alpha t)^2+\delta^2}= \frac{\hbar \alpha \delta}{2 \Omega^2
%(t)}.
%\end{eqnarray}
It is clear that the nature of the coupling in the two bases is
very different: the coupling between diabatic states is constant
in time, while the adiabatic states are significantly coupled
only near $t=0$, the off-diagonal terms in (\ref{Heff}) being inversely proportional to the square of the distance between the adiabatic levels $2\Omega(t)$. This enables one to simplify the description of the LZ transition in the adiabatic basis. As a useful approximation, we can replace the Hamiltonian (\ref{Heff}) in an interval of width $T_{\mathrm{LZ}}\sim \delta/\alpha$ around $t=0$ with one
with constant coefficients
\begin{equation}
\label{hsmall}
H_{\rm{ad}}(t) \simeq H_{\rm{ad}}(0) 
%-\Omega(0) \sigma_z + \Gamma(0) \sigma_y 
= - \delta \sigma_z +
\frac{\hbar\alpha}{2\delta} \sigma_y, \qquad (|t| \lesssim T_{\mathrm{LZ}}/2)
\end{equation}
and assume that outside this time interval the adiabatic
states evolve uncoupled, $H_{\rm{ad}}(t) \simeq -\alpha t \sigma_z$.

The afore-mentioned transition time $T_{\mathrm{LZ}}$ can be
fixed by imposing that the probability that the system, prepared in state $\ket{1(t)}$ at $t\to -\infty$, evolves into 
$\ket{2(t)}$ at $t\to +\infty$, be given by the Landau-Zener transition probability
\begin{equation}
\label{LZprob}
P_{\mathrm{LZ}}= \exp \left( -\frac{ \pi \delta^2 }{\hbar \alpha}
\right).
\end{equation}
Moreover, the $\sigma_z$ term in the Hamiltonian  (\ref{hsmall}) can be
neglected with good approximation provided
\begin{equation}
\frac{\Omega(0)}{\Gamma(0)} = \frac{2 \delta^2}{\hbar\alpha} = -
\frac{2}{\pi} \ln P_{\mathrm{LZ}}
\end{equation}
be sufficiently small. Within this approximation, the (unitary) evolution in the
interval $(-T_{\mathrm{LZ}}/2,T_{\mathrm{LZ}}/2)$ is governed by
\begin{equation}
\label{ueff}
U= \exp\left( - i \frac{\hbar\alpha }{\delta} \frac{T_{\mathrm{LZ}}}{2} \sigma_y 
\right) = \left(
\begin{array}{cc}
\cos\left(\hbar\alpha T_{\mathrm{LZ}}/2\delta \right) &
-\sin\left(\hbar\alpha T_{\mathrm{LZ}}/2\delta \right) \\
\sin\left(\hbar\alpha T_{\mathrm{LZ}}/2\delta \right) &
\cos\left(\hbar\alpha T_{\mathrm{LZ}}/2\delta \right)
\end{array} \right)
\end{equation}
with
\begin{equation}
\label{fp}
\sin^2 \left( \frac{\hbar\alpha}{2\delta} T_{\mathrm{LZ}} \right)
= P_{\mathrm{LZ}}.
\end{equation}
In the next section we will discuss how to apply this model for LZ
transitions to the problem of interband transitions in a
sinusoidal lattice.

\section{Interband tunneling in a lattice}

We are interested in describing the tunneling process from the
first to the second band of a Bose-Einstein condensate trapped in
an optical lattice. We assume that almost all atoms are in the condensate, so that the system
is described by a single-particle wave function $\psi(x,t)$ \cite{Stringari}. Moreover, let the condensate be dilute
enough so that the interaction between particles can be neglected.
This implies that the wave function of the system obeys a linear
Schr\"odinger equation.
 
The experimental condition is that of an accelerating
one-dimensional optical lattice, with constant acceleration $a$ \cite{Zenesini:2009}.
In the rest frame of the lattice, a particle of mass $m$ in the lattice is subjected to an external force $F=ma$ and a potential 
$(V/2)\cos (2\pi x/d_L)$, $V$ being the lattice depth and $d_L$ the lattice period (half wavelength  of the counterpropagating laser beams). The dynamics of the system depends on two
dimensionless parameters \cite{Tayebirad:2010}, related to lattice
depth and external force:
\begin{equation}
\label{rescaling} 
V_0=\frac{V}{E_{\mathrm{rec}}}, \quad
F_0=\frac{F d_L}{E_{\mathrm{rec}}}, \qquad \mathrm{ with } \quad
E_{\mathrm{rec}}=\frac{\hbar^2}{2m}\left( \frac{\pi}{d_L}
\right)^2 .
\end{equation}
In the adiabatic approximation, no transition takes place between bands (single-band approximation). This is consistent if
$Fd_L\lesssim V$, namely if  $F_0 \lesssim V_0$ in Eq.\
(\ref{rescaling}), and leads us to the two-level approximation outlined in the previous section.

We are interested in experimental setups in which the initial
state is highly peaked around a single quasimomentum value $k_0$,
that is, the width of the initial quasimomentum distribution is much smaller than the width $2\pi/d_L$ of the first Brillouin zone $\mathcal{B}$. Under this condition, in the adiabatic single-band approximation, the
average quasi-momentum evolves semiclassically \cite{Jones,Holthaus}, so that at time $t$
\begin{equation}\label{kclass}
k(t)=k_0+\frac{F t}{\hbar},
\end{equation}
with negligible spread in the distribution occurring during the
evolution. This yields Bloch oscillations in a tilted lattice with
a Bloch period
\begin{equation}\label{Tbloch}
T_{\rm B}=\frac{2\pi\hbar}{F
d_L}=\frac{\hbar}{E_{\mathrm{rec}}}\frac{2\pi}{F_0}.
\end{equation} The initial state analyzed here has a well
defined initial momentum (in $\mathcal{B}$), but can be
distributed among different bands. At the end of each Bloch
period, the amplitude in band $\alpha$ acquires the following
phase with respect to the amplitude in band $\beta$
\begin{equation}\label{Blochphase}
\Delta\varphi_{\alpha\beta}=  -\frac{2\pi}{F_0} \langle
E_{\alpha}(k)-E_{\beta}(k) \rangle ,
\end{equation}
where $\langle \dots\rangle$ denotes the average over $\mathcal{B}$ and $E_{\gamma}(k)$ is the energy of the state with quasimomentum $k$ in band $\gamma$ in units $E_{\mathrm{rec}}$. The relative phases acquired by states in different bands are at the origin of St\"uckelberg oscillations in the interband transition rates \cite{stuckosc}.

If the initial quasimomentum distribution is very peaked around
its central value, the interband transition can be analyzed as a
Landau-Zener tunneling, since in suitable parameter ranges the
transition from the first to the second band is concentrated at
the edges of the Brillouin zone, where an avoided crossing occurs.
In the Pisa experiment analyzed here \cite{LP}, this leads to an alternation of plateaus and
steep transition regions of the survival probability in the first band (see Fig.\ \ref{SurvivalProbability}). The
diabatic basis is represented by the momentum eigenstates, crossing at $\hbar\pi/d_L$ and coupled with strength $V/4$, while the adiabatic basis is represented by the quasimomentum eigenstates.
The LZ parameters in Eq.\ (\ref{LZprob}) are
\begin{equation}
\alpha= \frac{\pi\hbar F}{m d_L}, \qquad \delta= \frac{V}{4},
\end{equation}
yielding the transition probability (\ref{fp})
\begin{equation}\label{plzlatt}
P_{\mathrm{LZ}}^{(1,2)}(V_0,F_0)= \exp\left(-\frac{\pi^2 V_0^2}{32
F_0}\right).
\end{equation}
Even if the essential features of the transition are included in Eq.\ (\ref{plzlatt}), discrepancies can arise between the idealized case and the real physical situation. Indeed, the periodicity of the lattice implies that the afore-mentioned process occurs in a {\it finite} time, and that in
the initial and final states the adiabatic levels are not
infinitely separated. The corrections to the LZ transition
probability due to the finite duration of the process are
discussed in \cite{Holthaus,Tayebirad:2010}. Other corrections to
Eq.\ (\ref{plzlatt}) should be considered if the lattice is not
shallow. In this case, couplings to higher momentum states play an
important role and a two-level description is no longer a good
approximation.

Moreover, there is another kind of deviation from LZ, which is the
main object of our analysis. Since Eq.\ (\ref{plzlatt}) is
obtained under the hypothesis that only one of the two adiabatic
states is initially populated, it is no longer valid if both
states are populated. These deviations can be relevant even if one
of the initial populations is very close to zero, since they scale
as the {\it square root} of the smallest population, as will be
discussed in the following. These interference effects lead to
resonantly enhanced tunneling (RET): the transition probability is
enhanced by a large factor with respect to the LZ prediction if
the energy difference $Fd_L\Delta i$ between two potential wells
($d_L$ being the lattice spacing and $d_L\Delta i$ the distance
between the wells) matches the average band gap of the non-tilted
system.

The dynamics of interband tunneling can be schematized by
separately describing the transition at the avoided crossing and
the evolution far from the edges of the Brillouin zone. We will
assume that the transition between the first and the second band occurs in a time that is negligible with respect to the Bloch
time defined in (\ref{Tbloch}), namely $T_{\mathrm{LZ}}\ll T_B$.
We are thus assuming that the evolution inside the first band is adiabatic for all $k$, except for $k \simeq \pi/d_L$, when a
transition towards the state with the same quasi-momentum in the
second band is possible.

This transition can be effectively described by an evolution
operator corresponding to the one defined in Eq. (\ref{ueff}):
\begin{equation}
\tilde{U}=\left(\begin{array}{cc} s_{12} & -p_{12} \\ p_{12} &
s_{12}
\end{array}\right) ,
\end{equation}
with $s_{12}=\sqrt{P_{\mathrm{LZ}}(V_0,F_0)}$ and
$p_{12}=\sqrt{1-s_{12}^2}$. The operator $\tilde{U}$ acts on the
two-dimensional space spanned by $\{\ket{1},\ket{2}\}$, the states with the same quasimomentum in the first and the second band, respectively, that evolve according to (\ref{kclass}) with $k_0=0$.

The transition from the second to the third band can be
schematized as the loss of a fraction $1-s_{23}^2$ in the population of the second band towards a continuum, occurring at
the crossing around $k=0$. This assumption is justified for small values of $V_0$, such that a particle in the third (or higher)
band can be considered free. The survival amplitude $s_{23}$ can be determined by imposing that
its square be equal to $1-P_{\mathrm{LZ}}^{(2,3)}(V_0,F_0)$, where
$P_{\mathrm{LZ}}^{(2,3)}(V_0,F_0)$ the LZ transition probability
at the avoided crossing from the second to the third band
\begin{equation}
P_{\mathrm{LZ}}^{(2,3)}(V_0,F_0)= \exp\left(-\frac{\pi^2
V_0^4}{2^{14} F_0}\right).
\end{equation}
During each Bloch cycle separating two successive transitions, the relative phase between the second
and the first band amplitudes increases by the quantity (\ref{Blochphase}), that reads
\begin{equation}
\phi(V_0,F_0)= - \frac{2\pi}{F_0} \langle \Delta E(V_0) \rangle,
\label{phi12}
\end{equation}
where $\langle \Delta E \rangle$ is the energy difference (in
units $E_{\mathrm{rec}}$) between the second and the first band,
averaged over $\mathcal{B}$. The effects of
the dynamics in a time $T_{\rm B}$ from one transition to the next
one can thus be modelled in the basis $\{\ket{1},\ket{2}\}$ by an
effective non-unitary operator
\begin{equation}\label{opW}
W=\left(
\begin{array}{cc}
1 & 0 \\ 0 & s_{23}\mathrm{e}^{i\phi}
\end{array} \right) .
\end{equation}
The lack of unitarity is due to the two-level approximation adopted in our analysis and signals the flow of population out of the relevant two-dimensional subspace. Observe that the global evolution of the condensate is always coherent and does not involve any atomic losses.

By making use of this simplified model, we describe the time
evolution in the following way. At $t=0$ the condensate is in the first band, with quasi-momentum close to
$k=0$. As the lattice is accelerated, the quasi-momentum increases until it reaches $\pi/d_L$ at $t=T_{\rm B}/2$, where the operator $\tilde{U}$ comes into play and transfers part of the population
to the second band. The evolution from $T_{\rm B}/2$ to $3T_{\rm B}/2$ is summarized by the application of $W$. Then, the second
transition occurs, and part of the population in the second band
(minus losses towards the third band) can tunnel back to the first band due to the action of $\tilde{U}$, giving rise to interference effects. The same steps occur in the subsequent
transitions.

On a time span $T_{\rm B}$, the dynamics of the system is determined by the action of the non-unitary operator
\begin{equation}
\label{evored}
U=\tilde{U}W=\left( \begin{array}{cc} s_{12} &
-p_{12}s_{23}\mathrm{e}^{i\phi} \\ p_{12} & s_{12}
s_{23}\mathrm{e}^{i\phi}
\end{array}\right)
\end{equation}
in the basis $\{ \ket{1},\ket{2} \}$. The order of the two
operations is not relevant, since $W$ acts trivially on the
initial state $\ket{1}$ before the first transition.

\section{Transient and asymptotic dynamics}

%We now specialize the model outlined in the preceding section to the
%Pisa experimental setup \cite{Zenesini:2009,Tayebirad:2010}. 

We now look at the time evolution evolution implied by
the model outlined in the preceding section.
The
state of the system before the first transition is $\ket{1}$.
Immediately after the $n$-th transition, occurring at time
$t=T_{\rm B}(n+1/2)$, the state of the system is
\begin{equation}\label{evolved}
\ket{\Phi_n}=U^n\ket{1}.
\end{equation}
Let $\psi_1$ and $\psi_2$ be the normalized {\it non-orthogonal} eigenvectors of the matrix $U$ belonging to the eigenvalues $e_1$ and $e_2$, respectively.
The initial state
\begin{equation}\label{initial}
\ket{1}=c_1\ket{\psi_1}+c_2\ket{\psi_2},
\end{equation}
will evolve  at time $T_{\rm
B}(n+1/2)$ into
\begin{equation}
\label{evolution}
\ket{\Phi_n}=c_1 e_1^n \ket{\psi_1} + c_2 e_2^n \ket{\psi_2}.
\end{equation}
Due to the non-unitarity of $W$, the two eigenvalues are
smaller than unity. Let us assume that the eigenvalue $e_1$ is
larger in modulus than $e_2$, so that it eventually dominates in the iteration (\ref{evolution}). (This condition is necessarily verified for $0<s_{12},s_{23}<1$.) For $n$ sufficiently large, the evolution reaches an asymptotic regime, in which the state after the $n$-th
transition is determined only by the state after the previous one,
with a transition rate depending on the largest eigenvalue. Since
the survival probability in the first band can be defined as
$P_n=|\langle 1 | \Phi_n \rangle|^2$, in the asymptotic regime one
gets
\begin{equation}\label{gamma1}
P_n\simeq |e_1|^2 P_{n-1}.
\end{equation}
We define the asymptotic transition rate
\begin{equation}
\label{gamma2}
\gamma =-\ln \left( |e_1|^2 \right),
\end{equation}
and introduce an exponential function $P_{\rm Z}(t)$ that
coincides with the survival probability at the center
of the plateaus, at times $t=nT_{\rm B}$:
\begin{equation}\label{pzeta}
P_{\rm Z}(t)= Z \exp \left( -\gamma t \right). \label{expfit}
\end{equation}
This is the dashed line plotted in Fig.\ \ref{SurvivalProbability}. (Note the logarithmic scale on the ordinates.)

Observe that Eqs.\ (\ref{gamma1})-(\ref{pzeta}) are valid in the
asymptotic regime. Before reaching it, the ratio between $P_n$ and
$P_{n-1}$ in Eq. (28) depends on $n$ through a "time-dependent
decay rate" $\gamma_n$, as in
\begin{equation}
\label{gamman}
P_{n+1}=\mathrm{e}^{-\gamma_n} P_n.
\end{equation}
It is easy to prove that the succession $\gamma_n$ converges to the value $\gamma$ of Eq.\ (\ref{gamma2}), unless the
eigenvalues of $U$ have equal moduli. Indeed
\begin{equation}
\gamma_n= -\ln\frac{P_{n+1}}{P_n} = -\ln \left| \frac{ c_1
e_1^{n+1} \langle 1 \ket{\psi_1} + c_2 e_2^{n+1} \langle 1
\ket{\psi_2} }{ c_1 e_1^{n} \langle 1 \ket{\psi_1} + c_2 e_2^{n}
\langle 1 \ket{\psi_2} } \right|^2 ,
\end{equation}
which, by definition of $\gamma$ in Eq. (\ref{gamma2}), reads
\begin{equation}
\label{gammangamma}
\gamma_n = \gamma - 2\ln\left| \frac{ 1 + \frac{c_2 \langle 1
\ket{\psi_2}}{c_1 \langle 1 \ket{\psi_1}}
\left(\frac{e_2}{e_1}\right)^{n+1}  }{ 1 + \frac{c_2 \langle 1
\ket{\psi_2}}{c_1 \langle 1 \ket{\psi_1}}
\left(\frac{e_2}{e_1}\right)^{n} } \right| .
\end{equation}
Thus, if $|e_2|<|e_1|$, the argument in the logarithm approaches
one and $\gamma_n\to\gamma$.

\section{Wave function renormalization}

The parameter $Z$ in Eq.\ (\ref{expfit}) is in general different
from unity, due to the transient regime at the beginning of the
evolution. It represents the extrapolation of the
asymptotic exponential probability back at $t=0$. We derive below an expression for $Z$. In the asymptotic regime, the system evolution described by Eq.\ (\ref{evolution}) corresponds to an evolution operator applied
to an initial unnormalized vector $\ket{\Psi_0}=
c_1\ket{\psi_1}$:
\begin{equation}
\ket{\Phi_n} \simeq c_1 e_1^n\ket{\psi_1}=U^n\left(
c_1\ket{\psi_1}\right) = U^n \ket{\Psi_0}.
\end{equation}
The $Z$ parameter, representing the extrapolation of the
asymptotic behavior back to $t=0$, can be defined as the square
modulus of the projection of the fictitious initial vector
$\ket{\Psi_0}$, onto the actual initial state $\ket{1}$
\begin{equation}\label{Zdef}
Z\equiv |\bra{1}\Psi_0\rangle|^2=|c_1|^2 |\bra{1}\psi_1\rangle|^2,
\end{equation}
which corresponds to an extrapolated ``survival probability'' in the subspace spanned by $\ket{1}$, evaluated at the initial time. 
%$Z$ can be computed as a function of the independent parameters of the model, by explicitly diagonalizing $U$. 
In order to gain a qualitative understanding of the dependence of $Z$ on the phase difference $\phi$ acquired during a Bloch cycle, let us compare the first and second transitions. Let $P_0=1$ be the initial value of the survival probability in the first band. After the first transition, application of Eq.\ (\ref{evored}) yields the survival probability 
\begin{equation}
P_1=s_{12}^2 P_0\equiv \mathrm{e}^{-\gamma_0}P_0.
\end{equation}
At the second transition, the discrepancy with the LZ prediction
becomes manifest. In the parameter regime of small $V_0$ we
are considering, the ratio $s_{23}/s_{12}$ is very small \cite{LP} and we can apply a
first-order approximation, yielding
\begin{equation}
P_2\simeq (s_{12}^2 - 2 s_{23} p_{12}^2 \cos\phi) P_1\equiv
\mathrm{e}^{-\gamma_1}P_1.
\end{equation}
This clarifies that the plateaus in Fig.\ \ref{SurvivalProbability} are not equal.
If the phase is  $\phi=2\pi j$, with integer $j$, the
second transition is enhanced with respect to the first one. 
Thus, comparing with Eq.\ (27), local maxima in the transition rate are expected when $F_0(\phi=2\pi j) = \langle \Delta E(V_0) \rangle / j$. 

\begin{figure}
\begin{center}
\includegraphics[width=0.5\textwidth]{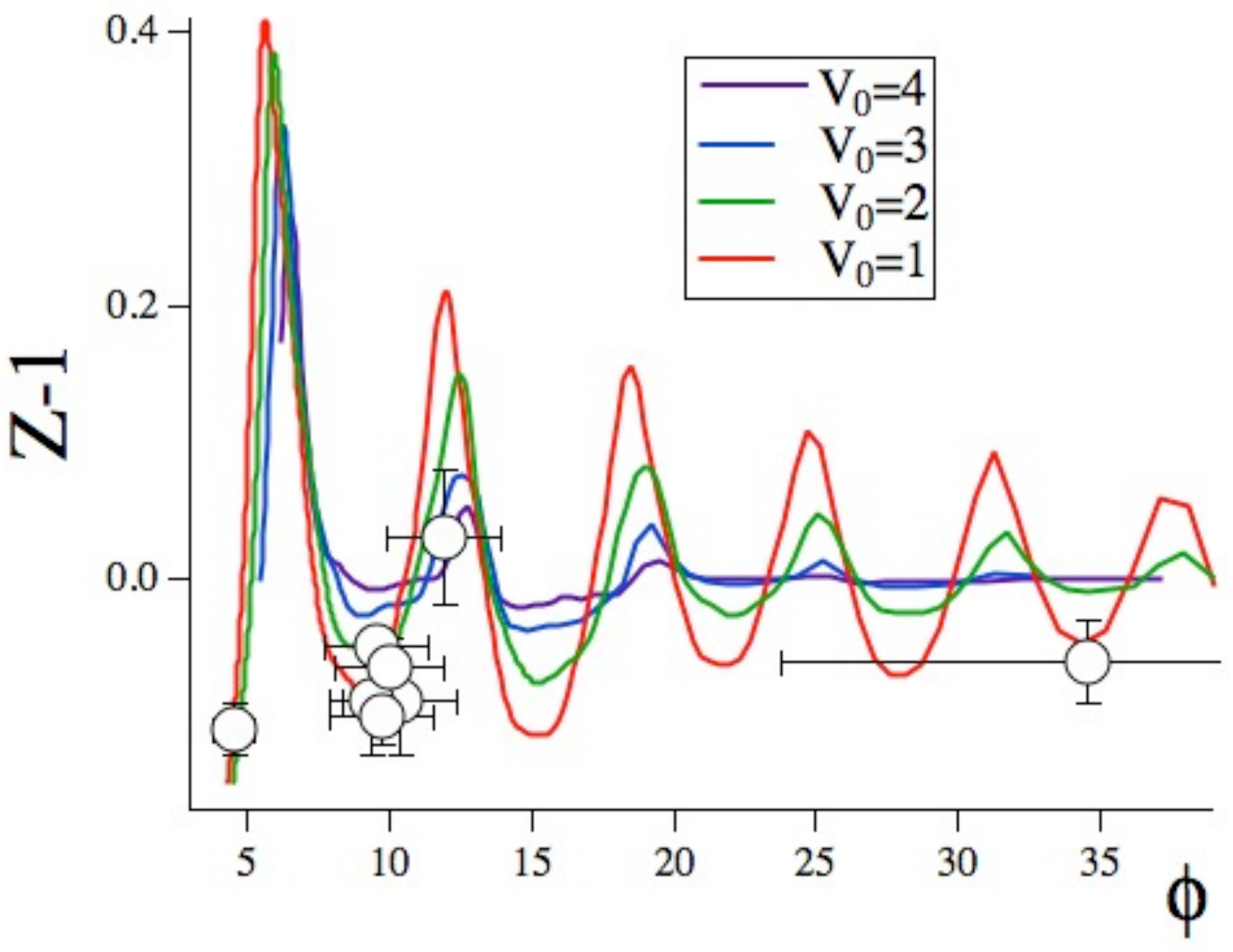}
\end{center}
\caption{Scaling plot of $Z-1$ vs.\ $\phi$ of Eq.\ (\ref{phi12}), derived from RET experimental results obtained in Ref.\ \cite{LP} (open circles) using a narrow atomic quasi-momentum distribution. The full lines are the theoretical predictions for $V_0=1,2,3,4$ (a smaller value of $V_0$ yielding wider oscillations).}
\label{RETScaling}
\end{figure}

This is the mechanism at the origin of wave function renormalization. A backwards extrapolation of the second step gives a rough estimate of the $Z$ parameter, which we call $Z_1$:
\begin{equation}
\label{Z1}
Z_1=\exp( \gamma_1-\gamma_0) \simeq 1+ 2 s_{23}
\left(\frac{p_{12}}{s_{12}}\right)^2 \cos\phi.
\end{equation}
Even if Eq.\ (\ref{Z1}) represents a rather crude approximation,
it is very useful in an experimental context, where only the first few steps in the Bloch cycles are
accessible. If the survival amplitude can be measured up to the
$N$-th transition, the $Z$ parameter can be approximated by
\begin{equation}
\label{ZN} 
Z_N= \exp\Big(N\gamma_N-\sum_{n=0}^{N-1} \gamma_n\Big) \to Z ,
\end{equation}
where the convergence to a fixed asymptotic value $Z$ is due to (\ref{gammangamma}). The convergence to $Z$ is typically very fast, and the first few cycles are already sufficient to obtain an excellent approximation. $Z$ is displayed in Fig. \ref{RETScaling} as a function of $\phi$.

\section{Conclusions}

In this article we studied Landau-Zener transitions and used them as a benchtest for the study of wave-function renormalization effects in quantum decay processes. By scrutinizing the features of the survival probability of the wave function that collectively describes an ultra-cold atomic cloud, we consistently defined $Z$ and extracted information on its behavior as a function of the experimental parameters. 

The value of $Z$ reflects the overlap between a discrete state and a continuum of states (to which the discrete state decays). Pictorially, one might say that $Z$ detects the overlap $Z=|\langle \psi_\mathrm{G} | \psi_\mathrm{in} \rangle|^2$ between the (generalized) decaying eigenfunction with complex energy eigenvalue,  the Gamow state $\psi_\mathrm{G}$, and the initial (physical) state $\psi_\mathrm{in}$. The case $Z<1$ does not surprise, being in accord with the familiar K\"all\'en-Lehmann paradigm \cite{KL} for stable asymptotic states. The situation $Z>1$ is more curious, and is a consequence of the fact that $\psi_\mathrm{G}$ does not live in the Hilbert space, its norm being infinite \cite{Gamow,Bohm,Gadella}, and its overlap $Z$ with the initial state  can exceed 1.

 It is remarkable that a quantity like $Z$ can be directly measured and that its deviations from unity yields directly measurable consequences. In addition, as the experimental parameters are varied, $Z$ can take values that can be smaller or larger than unity.

\section*{Acknowledgments}
We thank G. Florio, H. Lignier, N. L\"orch, R. Mannella and S. Wimberger for many interesting discussions. The experiment discussed in this article was done in Pisa. The numerical analyses in Figs. \ref{SurvivalProbability} and \ref{RETScaling} were performed by N. L\"orch and S. Wimberger.

\end{document}